\documentclass[aps,prd,twocolumn,groupedaddress,showpacs]{revtex4}
\usepackage{graphicx}
\usepackage{amsmath}
\usepackage{textcomp}

\usepackage{dcolumn}
\usepackage{bm}
\usepackage[latin1]{inputenc}
\usepackage{amssymb}
\usepackage{amsfonts}
\usepackage{lineno}
\usepackage{url}
\usepackage{array}
\usepackage{verbatim}
\usepackage{subfigure}
\usepackage{color}
\usepackage{lscape}

\begin{document}


\title{Limits on the Axial Coupling Constant of New Light Bosons}

\author{Florian M. Piegsa}
\affiliation{ETH Z\"urich, Institute for Particle Physics, CH-8093 Z\"urich, Switzerland}
\email{florian.piegsa@phys.ethz.ch}

\author{Guillaume Pignol}
\affiliation{LPSC, Universit\'e Joseph Fourier, CNRS/IN2P3, INPG, F-38000 Grenoble, France}
\email{guillaume.pignol@lpsc.in2p3.fr}

\date{\today}

\begin{abstract}
We report on a neutron particle physics experiment, which provides for the first time an upper limit on the strength of an axial coupling constant for a new light spin 1 boson in the millimeter range. Such a new boson would mediate a new force between ordinary fermions, like neutrons and protons. The experiment was set up at the cold neutron reflectometer Narziss at the Paul Scherrer Institute and uses Ramsey's technique of separated oscillating fields to search for a pseudo-magnetic neutron spin precession induced by this new interaction. For the axial coupling constant $g_{\text{A}}^2$ an upper limit of $6 \times 10^{-13}$ (95\% C.L.) was determined for an interaction range of 1 mm.

\end{abstract}

\pacs{03.75.Be, 07.55.Ge, 14.70.Pw, 34.20.Cf}






\maketitle



The Standard Model (SM) of particle physics explains the interaction between matter particles in terms of exchange of bosons. It is believed that the SM corresponds to the low energy limit of a more fundamental theory, that should ultimately also describe gravity in the same framework. Various extensions of the SM based on new ideas, such as superstring theory, are being developed in this direction; they all generically predict new particles and thus new interactions beyond the SM. A new boson with mass $M$ would induce an interaction between ordinary particles with an interaction range $\lambda_c = \hbar/Mc$. The Large Hadron Collider can probe new bosons with a mass up to $1~\text{TeV}/c^2$, provided they are strongly coupled with matter. Lighter bosons could have escaped detection up to now if they are weakly coupled. In the latter case one does not search for the boson itself but for the additional interaction the boson mediates between ordinary particles. All possible interactions have recently been classified \cite{1}. This letter reports the first result from a table-top experiment designed to probe spin-dependent interactions in the millimeter range, induced by a vector boson with a mass of about $10^{-4}$~eV. \\
In the framework of relativistic quantum field theory, the interaction between a nucleon $\psi$ and a new vector (spin 1) boson $X^\mu$ is generically characterized by the Lagrange function $\mathcal{L} = \overline{\psi} (g_{\text{V}} \gamma^\mu + g_{\text{A}} \gamma^\mu \gamma^5) \psi X_\mu$ which is parametrized by a vector coupling constant $g_V$ and an axial-vector coupling constant $g_A$. For the sake of simplicity we assume identical coupling to protons and neutrons and no coupling to electrons. The vector term induces a spin-independent repulsive force with a strength proportional to $g_V^2$. Numerous experiments are searching for such a fifth force with different techniques to probe ranges from subatomic to astronomical distances \cite{2,3,4,5,6,30}. At the millimeter range, the best limit is extracted from the Seattle torsion pendulum experiment \cite{5}, looking at tiny deviations from Newton's law of gravity acting on macroscopic bodies, with the present best upper limit of $g_{\text{V}}^2 < 5 \times 10^{-40}$. The axial-vector term induces a new spin-dependent interaction and can only be probed using spin polarized particles. One way to access this term is to look at a dipole-dipole type of interaction, as done in a recent experiment performed at Princeton \cite{12}. It uses a $^3$He spin source and a K-$^3$He co-magnetometer probe and reports the limit:
\begin{equation}
g_{\text{A}}^2 e^{-r/\lambda_{\text{c}}} < 1.5 \times 10^{-40}
\end{equation}
with $r=0.5$~m being the distance between the $^3$He spin source and the K-$^3$He co-magnetometer. \\
Another way to probe the axial-vector term is to look at the induced spin-velocity interaction between a particle considered as a source and another polarized probe particle. The corresponding interaction potential, in the non-relativistic limit, can be written as: 
\begin{equation}
\label{Vpoint}
V_{\text{Axial}}^{\rm{point}}(r) = 
\frac{g_{\text{A}}^2}{16 \pi} \, \frac{(\hbar c)^2}{m c^2} \, \vec{\sigma} \cdot \left( \frac{\vec{v}}{c} \times \frac{\vec{r}}{r} \right) \, \left( \frac{1}{\lambda_{\text{c}}} + \frac{1}{r}  \right) \,  \frac{e^{-r/\lambda_{\text{c}}}}{r}
\end{equation}
where $\vec{\sigma}$ is the spin and $m$ is the mass of the probe particle, $r$ is the distance and $\vec{v}$ is the relative velocity between the source and the probe, respectively. This potential corresponds to $f_{\bot}$ in the notation of \cite{1}. The fact that the interaction does not depend on the source particle's spin allows to use an unpolarized macroscopic object as a source.
\begin{figure}
\includegraphics[width=0.45\textwidth]{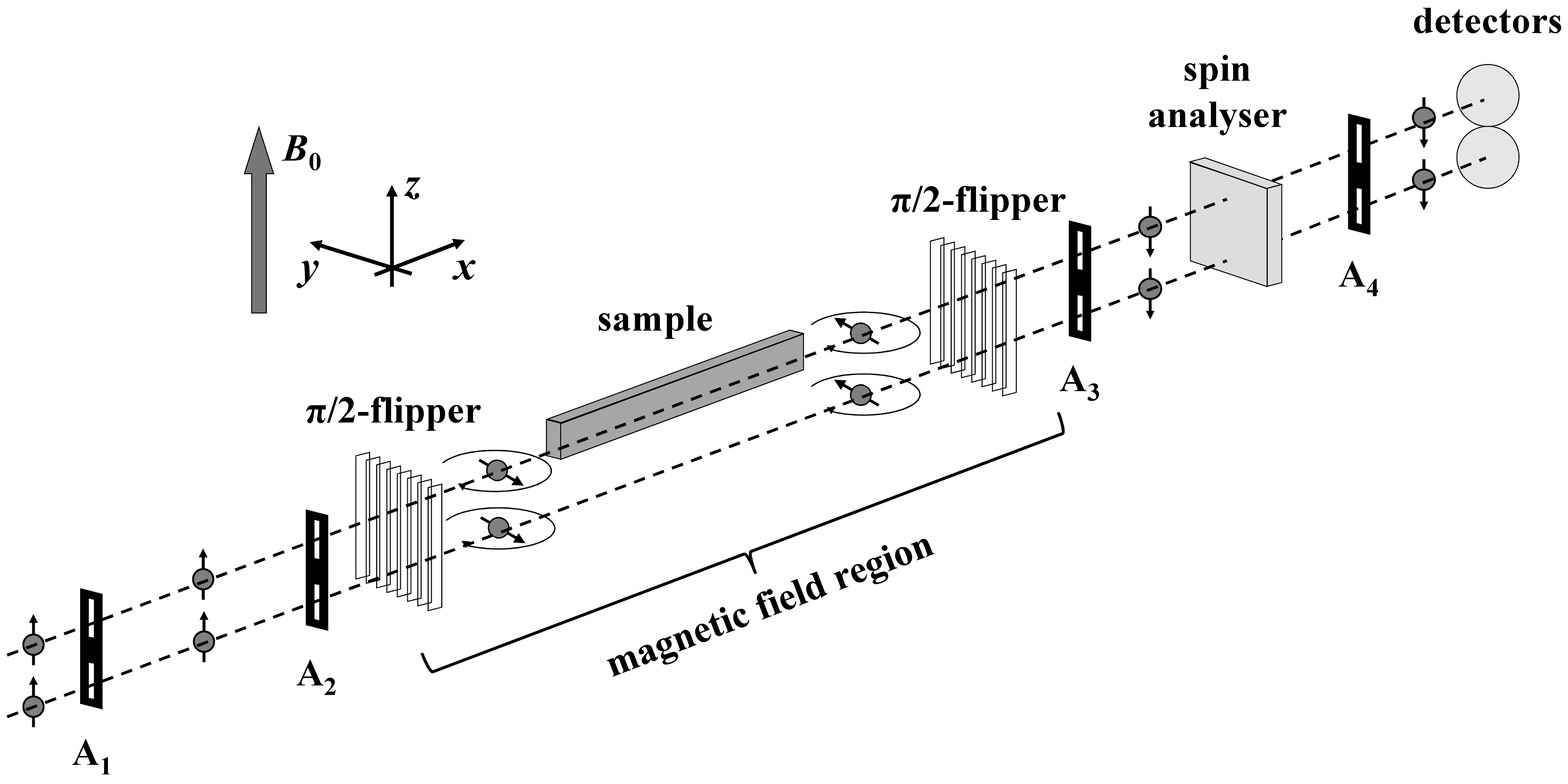}
\caption{Schematic of the experimental setup to probe the axial interaction between polarized neutrons and a macroscopic sample plate as a function of the distance $\Delta y$. The polarized neutron beam is separated into two beams by several apertures $A_{1-4}$. The sample is located between the two $\pi/2$ spin-flippers where the neutron spins precess in the $x$-$y$-plane perpendicular to the static magnetic field $B_0$.  }
\label{fig:setup}
\end{figure} \\

\begin{figure}
	\includegraphics[width=0.35\textwidth]{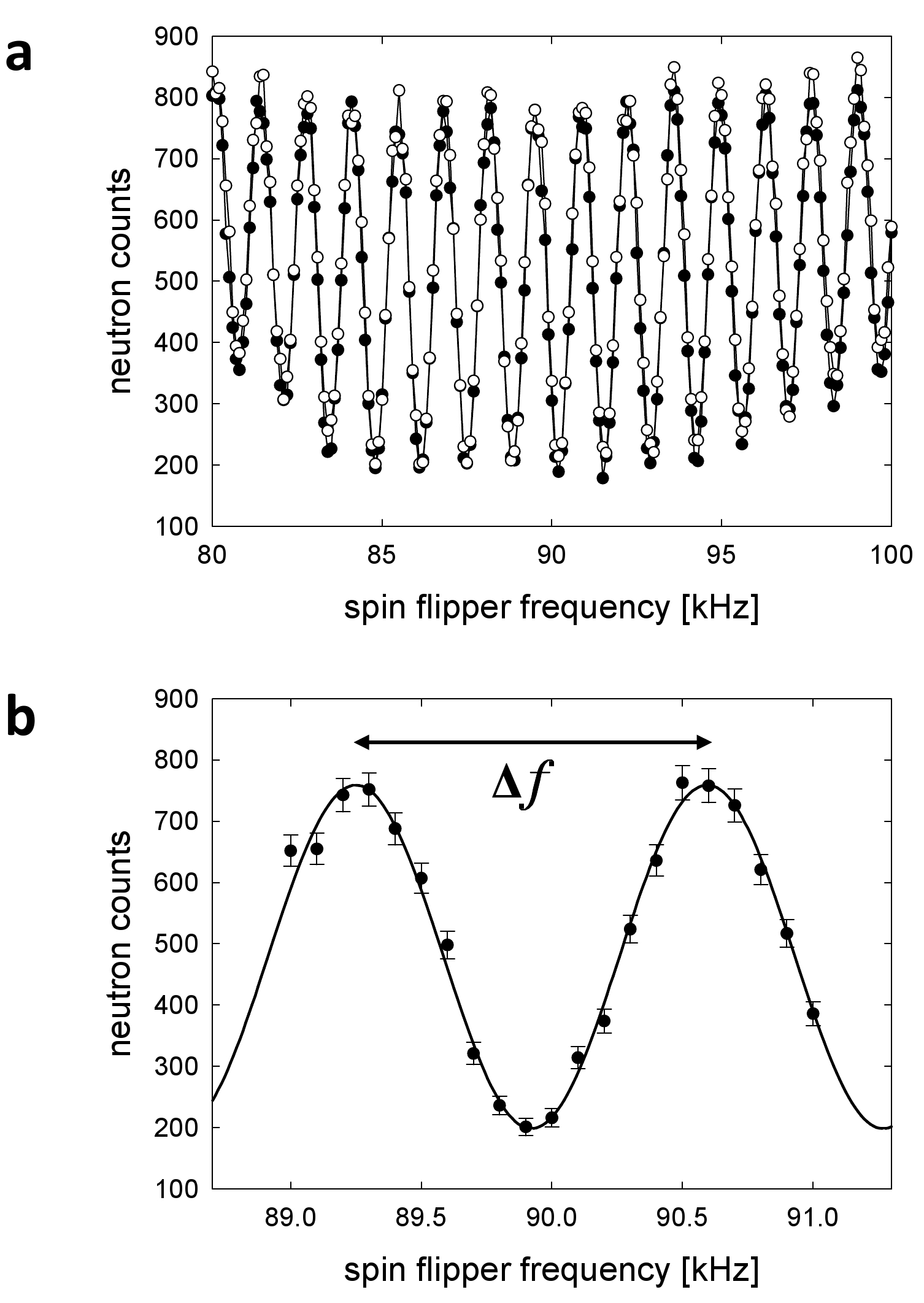}
	\caption{(a) Ramsey oscillation pattern of the sample beam (\textbullet) and the reference beam (\textopenbullet) close to the neutron Larmor resonance of about 90 kHz. (b) A reduced Ramsey pattern consisting only of 21 data points, measured in a total time of approximately 5 minutes at a proton current of 1.35 mA on the SINQ spallation target.}
	\label{fig:ramseys}
\end{figure}

Here, we describe how this exotic interaction is probed for the first time employing a beam of polarized slow neutrons as a probe and a macroscopic plate as a source, as proposed in \cite{19}. Near the surface of the plate the potentials (\ref{Vpoint}) induced by the individual nucleons in the bulk add up coherently. This therefore creates an effective pseudomagnetic field, directed orthogonal to both the neutron direction of flight and the surface normal of the sample. In the configuration presented in Fig. \ref{fig:setup}, the neutron velocity is directed along $x$ and the pseudo-magnetic field would be directed along $z$. If the neutron beam is initially polarized in the $x$-$y$-plane, the pseudo-magnetic field will induce a spin precession around the $z$-axis of
\begin{equation}
\varphi = l \ \frac{g_{\text{A}}^2}{4} \ N \ \frac{\hbar c}{m_{\text{n}} c^2} \ \lambda_{\text{c}} \ e^{-\Delta y/\lambda_{\text{c}}}
\label{equ:B_AA}
\end{equation}
where $\Delta y > 0$ is the distance between the neutron beam and the sample surface, $l$ is the length of the sample in $x$-direction, $N$ is the nucleon number density of the sample and $m_{\text{n}}$ is the neutron mass. \\
A very accurate method to measure such a neutron spin precession is Ramsey's technique of separated oscillating fields \cite{13}. In Fig. \ref{fig:setup} a schematic drawing of the utilized Ramsey setup is depicted, which was installed at the polarized neutron reflectometer Narziss at the spallation source SINQ at the Paul Scherrer Institute. The neutron beam with a de Broglie wavelength of 0.5 nm and a velocity spread of 1.5\% (FWHM) is polarized in the $z$-direction. It is shaped and separated by means of the apertures A$_1$ and A$_2$ into two beams each with a height of 10 mm. The distance in $z$-direction between the beams is 20 mm from center to center. The width of the upper beam was measured to be $(0.46 \pm 0.02)$~mm (FWHM). It passes by the sample in a distance $\Delta y$ and the lower beam, passing at a distance much larger than $\Delta y$, serves as a reference (\emph{two beam method}). Along the beam path two phase-locked radio frequency (rf) $\pi/2$ spin-flip coils, with a thickness of 10 mm and a distance to each other of 570~mm, are placed in a magnet which provides a static field $B_0$ of approximately 3 mT in $z$-direction. The field is continuously monitored using a 3D Hall probe which is installed outside of the homogeneous field region in the vicinity of the magnet in a field of 2.3 mT. The investigated sample is placed between the two $\pi/2$ spin-flip coils, which produce linear oscillating fields in $y$-direction. The spins of the neutrons are analyzed using a polarizing supermirror and are finally detected using two $^3$He gas detectors. The number of detected neutrons is normalized by means of a fission chamber monitor detector placed in the incident neutron beam. The additional apertures behind the sample (A$_3$ and A$_4$) avoid detecting neutrons which are accidentially scattered in air or on the sample surface. \\
A so-called Ramsey pattern is obtained by measuring the detector count rate as a function of the spin-flipper frequency close to the neutron Larmor resonance, as presented in Fig. \ref{fig:ramseys}a \cite{15,16,17}. Any additional precession of the neutron spins due to the sample will cause a corresponding phase shift of the Ramsey pattern of the sample beam, while the phase of the Ramsey pattern of the reference beam will stay unchanged. Instead of measuring a complete Ramsey pattern, only a set of 21 data points is taken to reduce measuring time, as shown in Fig. \ref{fig:ramseys}b. The sinusoidal fit to the data points provides a fit-accuracy of the phase of about $\pm1.4$°. The same fit yields a Ramsey oscillation frequency $\Delta f =(1353 \pm 12)$~Hz, which is in good agreement with the expected theoretical value of 1358~Hz determined using Eq. (4) in \cite{15}. 
\begin{figure}
	\centering
		\includegraphics[width=0.45\textwidth]{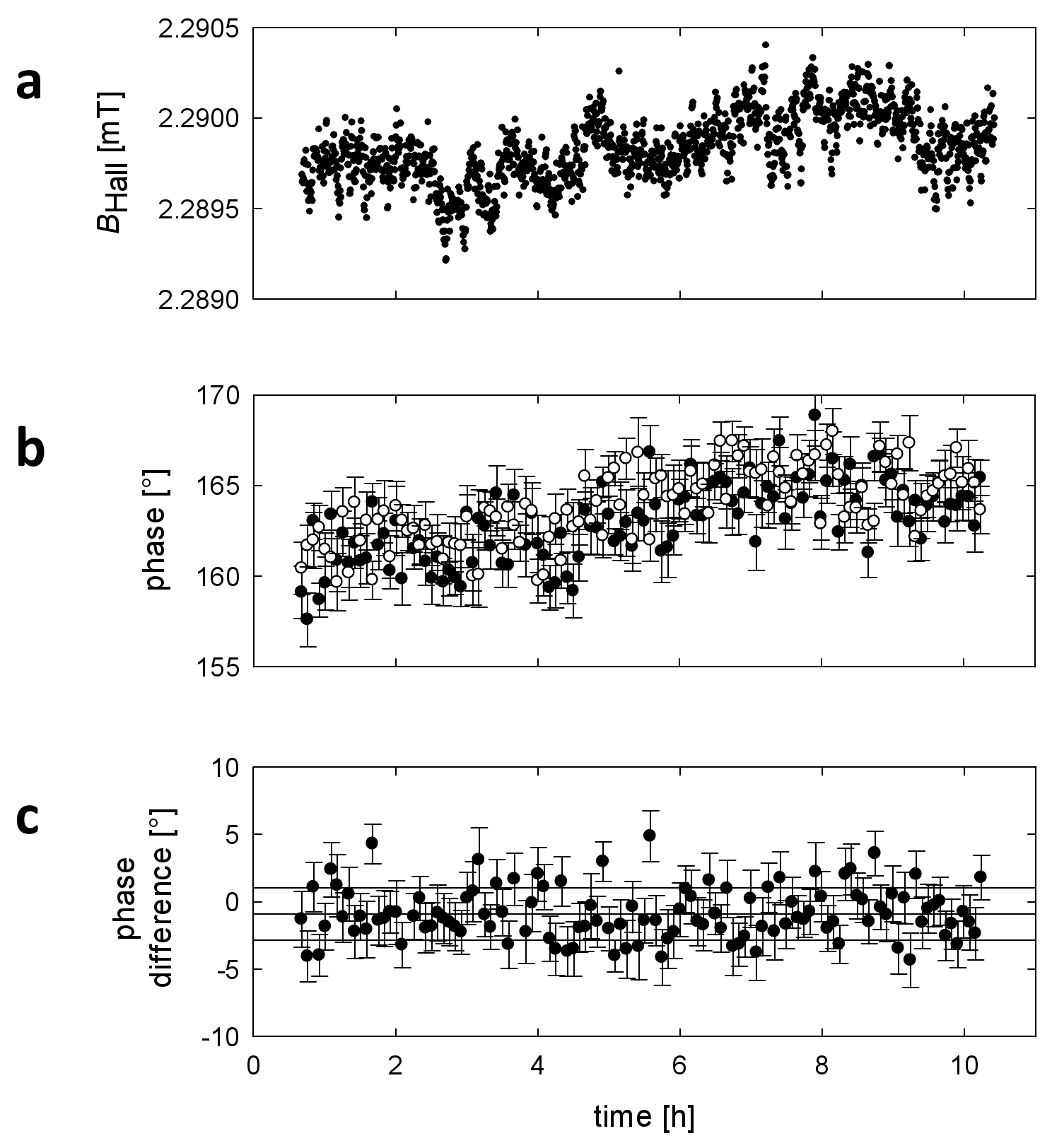}
	\caption{Stability test of the Ramsey apparatus during approximately 10 hours. (a) Time histogram of the field magnitude seen by the 3D Hall probe in the vicinity of the $B_0$-magnet. (b) Individual phases of the Ramsey patterns of the sample (\textbullet) and reference beam (\textopenbullet). (c) Phase difference of the two beams. The horizontal lines indicate the mean constant offset $(-0.9 \pm 0.2)$° and the standard deviation $\pm1.9$°. The average individual error is also $\pm1.9$°. }
	\label{fig:stability}
\end{figure}\\
The stability of the apparatus was tested by repeatedly performing Ramsey scans for almost 10 hours without a sample. The result is presented in Fig. \ref{fig:stability} and demonstrates that although the phases of the individual beams might drift slightly (probably due to magnetic field and thermal drifts), the phase difference remains constant and scatters purely statistically. This important advantage of the two beam method, i.e. allowing to compensate for inadvertent drifts, is mandatory to perform a high-accuracy measurement and was already demonstrated previously in \cite{14,15}. \\

\begin{figure}
	\centering
		\includegraphics[width=0.43\textwidth]{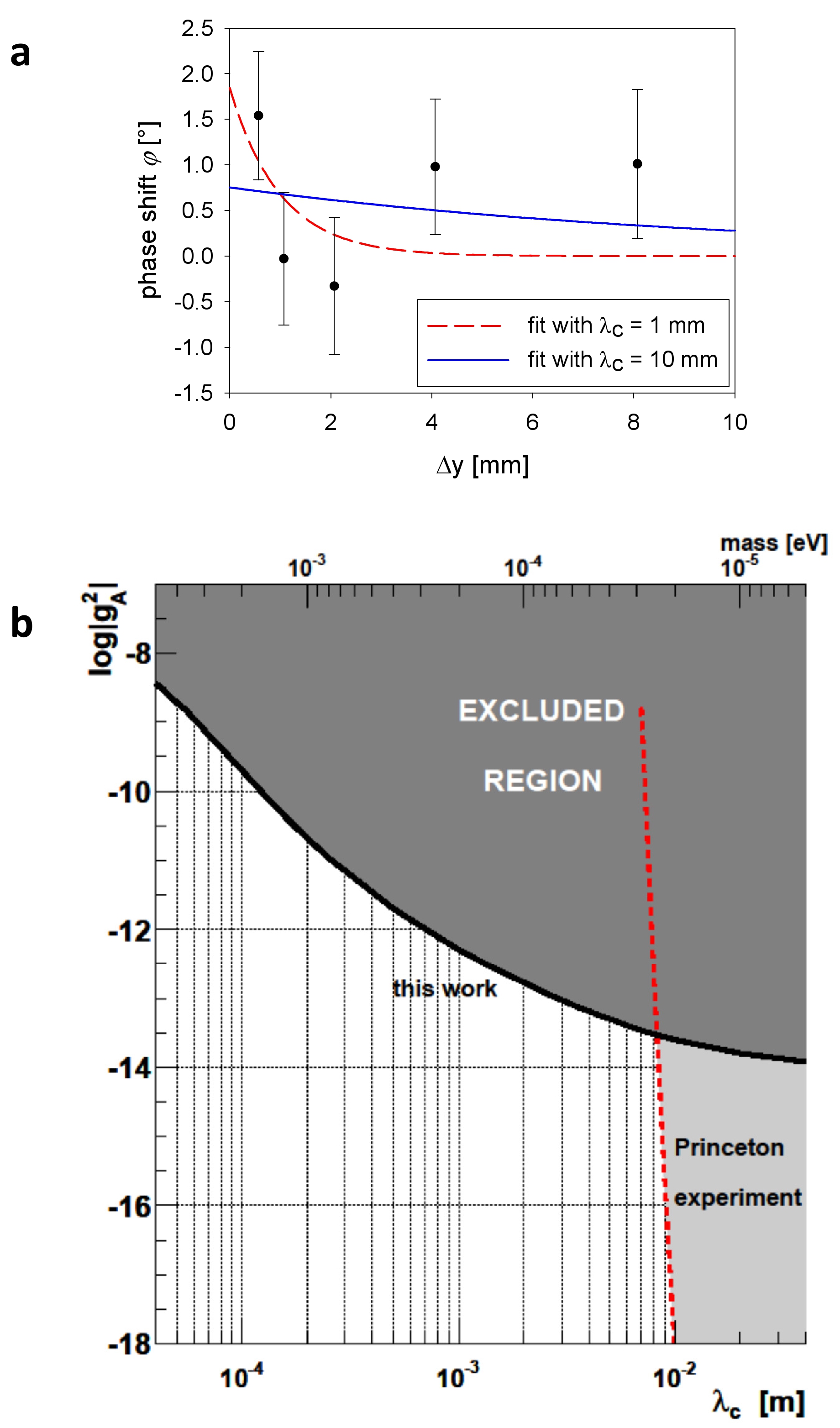}
	\caption{(a) Phase shift due to the sample as a function of the distance $\Delta y$. The data has been corrected for possible drifts using the reference beam. Two exponential fits to the data points with fixed $\lambda_{\text{c}}$ are presented. (b) Exclusion plot for the axial-axial coupling $g_{\text{A}}^2$ (95\% C.L.) as a function of the interaction range $\lambda_{\text{c}}$ and the mass $M$ of a new boson, respectively. The almost vertical dashed line represents the border of the exclusion region deduced from the experimental results by the Princeton group \cite{12}.}
	\label{fig:exclusion}
\end{figure}

As a sample, a 19 mm thick and 480 mm long copper plate was used (nucleon number density: $N\approx5.3\times10^{24}$~cm$^{-3}$). The copper plate surfaces were machined to achieve a roughness and error in parallelism of less than 50~$\mu$m and the sample was aligned parallel with respect to the neutron beam with an accuracy of better than 0.007°. The reproducibility of the sample positioning in $y$-direction was measured to be about $\pm 1$~$\mu$m and the position of the sample surface was determined with an accuracy of $\pm 10$~$\mu$m. This yields an overall error for $\Delta y$ of approximately $\pm 80$~$\mu$m, which is much smaller than the width of the neutron beam. \\
One measurement cycle consists of five Ramsey scans taken at $\Delta y$ equal to 0.57 mm, 1.07 mm, 2.07 mm, 4.07 mm and 8.07 mm. 8 successive cycles with and 20 cycles without the sample were performed, in a total time of 12 hours. The combined results are summarized in Fig. \ref{fig:exclusion}a, which shows the phase shift due to the sample as a function of the sample-to-beam distance.
By fitting the exponential decay function $\varphi(\Delta y)=\varphi_0\cdot e^{-\Delta y / \lambda_{\text{c}}}$ with fixed $\lambda_{\text{c}}$ to the data points, one obtains the parameter $\varphi_0$ with its uncertainty $\delta \varphi_0$ (e.g. for $\lambda_{\text{c}}=1$~mm: $\varphi_0=(1.84 \pm 1.27)$° and $\lambda_{\text{c}}=10$~mm: $\varphi_0=(0.75 \pm 0.48)$°). From that one can determine upper limits of the axial-axial coupling $g_{\text{A}}^2$ as a function of $\lambda_{\text{c}}$ at 95\% confidence level, using
\begin{equation}
g_A^2 < \frac{4 m_{\text{n}} c^2}{l N \hbar c} \cdot \frac{1}{\lambda_{\text{c}}} \cdot (\left| \varphi_0 \right| + 1.96 \cdot \delta\varphi)
\label{equ:gaa2}
\end{equation}
This leads to the exclusion plot presented in Fig. \ref{fig:exclusion}b.  It should be noted, that for $\lambda_{\text{c}} < 0.5$~mm the size of the neutron beam was taken into account by performing a convolution over the beam distribution. 
\\






Finally, we consider possible systematic effects which could disguise a real or produce a false phase shift of the Ramsey signal: 
\begin{itemize}
  \item One systematic effect might occur due to residual magnetic impurities at the surface of the sample. This was investigated by performing a fluxgate (Hall-probe) scan of the magnetic field in $z$-direction ($y$-direction) close to the surface of the copper plate, at a distance of 4 mm (2.5 mm). The scans did not show any field inhomogeneity larger than 10 nT peak-to-peak for the fluxgate and 100 nT peak-to-peak for the Hall probe, respectively \cite{20}. A numerical calculation  shows that a single dipole at the surface, consistent with the field measurements, cannot induce a spin rotation larger than 0.02°.
	\item The diamagnetic property of copper ($\chi = - 10^{-5}$) induces in the worst case, i.e. the neutron beam crosses the entire sample, a phase shift of -0.2° in a field of 3 mT \cite{18}. Since in the here presented experiment the neutrons do not penetrate the sample, the phase precession is much smaller and we can neglect this effect.
	\item The sample could influence the resonance condition of the $\pi/2$ spin-flip coils, which might produce a deviation of the relative phase of the locked rf signals. From off-line tests in which the rf signals of the spin flippers were recorded with and without sample using two pick-up coils, an upper limit of 0.1° can be stated.
	\item The largest field in the experimental area is produced by the sample magnet. All other sources of magnetic fields are further away and smaller. Hence, if $B_0$ drifts it could cause a change of the field gradient, which could yield a relative phase shift between the two beams. This effect is enormously suppressed due to the two beam method and is smaller than 0.1° for a field drift of 4~$\mu$T, which is large considering the excellent stability of $B_0$ presented in Fig. \ref{fig:stability}a.
\end{itemize}
Summarizing, these possible sources for systematic errors are much smaller than the obtained statistical accuracy of about $\pm 0.75$° (compare Fig. \ref{fig:exclusion}a) and, thus, can be neglected. However, if one increases the measuring time up to a week, chooses a broader neutron velocity spectrum of about 10\% and performs the experiment at a more intense neutron source (assuming a neutron flux increase by 10 to 100), the statistical precision could be improved with moderate effort by a factor 10 to 30. \\



In conclusion, the search for a new spin-dependent fundamental interaction acting on nucleons has recently attracted increasing interest of a growing community of researchers. The case of a scalar (spin 0) boson, mediating an \emph{Axion-like} interaction is actively investigated using atoms \cite{8}, ultracold neutrons \cite{7,10,22} and polarized $^3$He \cite{9}. We have presented a neutron Ramsey experiment to probe for the first time the case of a vector (spin 1) boson. This experiment yields an upper limit of $g_{\text{A}}^2 < 6\times 10^{-13}$ for the axial-axial coupling at $\lambda_{\text{c}} = 1$ mm and represents, to our knowledge, so far the best method to probe spin-velocity interactions in the millimeter range. \\




We gratefully acknowledge the help by the Laboratory for Developments and Methods of the Paul Scherrer Institute (B.~van~den~Brandt, P.~Hautle, P.~Schurter, M.~K\"onnecke and U.~Filges) and the "D-PHYS Zentralwerkstatt" of ETH~Z\"urich. Further, we thank C.~Schanzer and M.~Schneider from SwissNeutronics and K.~Kirch and O.~Zimmer for their support and many fruitful discussions. \\
This work was performed at the Swiss Spallation Neutron Source at the Paul Scherrer Institute, Villigen, Switzerland.

\end{document}